\documentclass{aastex}

\makeatletter
  
  \@addtoreset{equation}{section}
\makeatother

\slugcomment{The Astrophysical Journal, to be submitted}

\shortauthors{Itoh and Nozawa}
\shorttitle{CORRECTIONS TO THE S-Z EFFECT FOR CLUSTERS OF GALAXIES V}

\begin{document}

\title{RELATIVISTIC CORRECTIONS TO THE SUNYAEV-ZELDOVICH EFFECT FOR CLUSTERS OF GALAXIES. V. NUMERICAL RESULTS FOR HIGH ELECTRON TEMPERATURES}

\author{NAOKI ITOH}
\affil{Department of Physics, Sophia University, 7-1 Kioi-cho, Chiyoda-ku, Tokyo, \\
102-8554, Japan; n\_itoh@sophia.ac.jp}

\and

\author{SATOSHI NOZAWA}
\affil{Josai Junior College for Women, 1-1 Keyakidai, Sakado-shi, Saitama, \\
350-0295, Japan; snozawa@josai.ac.jp}

\begin{abstract}
  We present an accurate numerical table for the relativistic corrections to the thermal Sunyaev-Zeldovich effect for clusters of galaxies.  The numerical results for the relativistic corrections have been obtained by numerical integration of the collision term of the Boltzmann equation.  The numerical table is provided for the ranges $0.002 \leq \theta_{e} \leq 0.100$ and $0 \leq X \leq 20$, where $\theta_{e} \equiv k_{B}T_{e}/m_{e}c^{2}$, $X \equiv \hbar \omega/k_{B}T_{0}$, $T_{e}$ is the electron temperature, $\omega$ is the angular frequency of the photon, and $T_{0}$ is the temperature of the cosmic microwave background radiation.  We also present an accurate analytic fitting formula that reproduces the numerical results with high precision.
\end{abstract}

\keywords{cosmic microwave background --- cosmology: theory --- galaxies: clusters: general --- radiation mechanisms: thermal --- relativity}

\section{INTRODUCTION}

  Compton scattering of the cosmic microwave background (CMB) radiation by hot intracluster gas --- the Sunyaev-Zeldovich effect (Zeldovich \& Sunyaev 1969; Sunyaev \& Zeldovich 1972, 1980a, 1980b, 1981) --- provides a useful method for the studies of cosmology (see recent excellent reviews: Birkinshaw 1999; Carlstrom, Holder, \& Reese 2002).  The original Sunyaev-Zeldovich formula has been derived from a kinetic equation for the photon distribution function taking into account the Compton scattering by electrons: the Kompaneets equation (Kompaneets 1957; Weymann 1965).  The original Kompaneets equation has been derived with a nonrelativistic approximation for the electron.  However, recent X-ray observations have revealed the existence of many high-temperature galaxy clusters (David et al. 1993; Arnaud et al. 1994; Markevitch et al. 1994; Markevitch et al. 1996; Holzapfel et al. 1997; Mushotzky \& Scharf 1997; Markevitch 1998; Allen, Ettori, \& Fabian 2001; Schmidt, Allen, \& Fabian 2001).  In particular, Tucker et al. (1998) reported the discovery of a galaxy cluster with the electron temperature $k_{B} T_{e} = 17.4 \pm 2.5$ keV.  Rephaeli and his collaborator (Rephaeli 1995; Rephaeli \& Yankovitch 1997) have emphasized the need to take into account the relativistic corrections to the Sunyaev-Zeldovich effect for clusters of galaxies.

  In recent years remarkable progress has been achieved in the theoretical studies of the relativistic corrections to the Sunyaev-Zeldovich effects for clusters of galaxies.  Stebbins (1997) generalized the Kompaneets equation.  Itoh, Kohyama, \& Nozawa (1998) have adopted a relativistically covariant formalism to describe the Compton scattering process (Berestetskii, Lifshitz, \& Pitaevskii 1982; Buchler \& Yueh 1976), thereby obtaining higher-order relativistic corrections to the thermal Sunyaev-Zeldovich effect in the form of the Fokker-Planck expansion.  In their derivation, the scheme to conserve the photon number at every stage of the expansion which has been proposed by Challinor \& Lasenby (1998) played an essential role.  The results of Challinor \& Lasenby (1998) are in agreement with those of Itoh et al. (1998).  The latter results include higher-order expansions.  Itoh et al. (1998) have also calculated the collision integral of the Boltzmann equation numerically and have compared the results with those obtained by the Fokker-Planck expansion method.  They have confirmed that the Fokker-Planck expansion method gives an excellent result for $k_{B}T_{e} \leq 15$keV, where $T_{e}$ is the electron temperature.  For $k_{B}T_{e} \geq 15$keV, however, the Fokker-Planck expansion results show nonnegligible deviations from the results obtained by the numerical integration of the collision term of the Boltzmann equation.

  Nozawa, Itoh, \& Kohyama (1998b) have extended their method to the case where the galaxy cluster is moving with a peculiar velocity with respect to CMB.  They have thereby obtained the relativistic corrections to the kinematical Sunyaev-Zeldovich effect.  Challinor \& Lasenby (1999) have confirmed the correctness of the result obtained by Nozawa et al. (1998b).  Sazonov \& Sunyaev (1998a, b) have calculated the kinematical Sunyaev-Zeldovich effect by a different method.  Their results are in agreement with those of Nozawa et al. (1998b).  The latter authors have given the results of the higher-order expansions.

  Itoh, Nozawa, \& Kohyama (2000a) have also applied their method to the calculation of the relativistic corrections to the polarization Sunyaev-Zeldovich effect (Sunyaev \& Zeldovich 1980b, 1981).  They have thereby confirmed the result of Challinor, Ford, \& Lasenby (1999) which has been obtained with a completely different method.  Recent works on the polarization Sunyaev-Zeldovich effect include Audit \& Simons (1998), Hansen \& Lilje (1999), and Sazonov \& Sunyaev (1999).

  In the present paper we address ourselves to the numerical calculation of the relativistic corrections to the thermal Sunyaev-Zeldovich effect.  As stated above, Itoh et al. (1998) have carried out the numerical integration of the collision term of the Boltzmann equation.  This method produces the exact results without the power series expansion approximation.  In view of the discovery of an extremely high temperature galaxy cluster with $k_{B}T_{e} = 17.4 \pm 2.5$keV (Tucker et al. 1998), it would be extremely useful to present the results of the numerical integration of the collision term of the Boltzmann equation.
  Sazonov \& Sunyaev (1998a, b) have reported the results of the Monte Carlo calculations on the relativistic corrections to the Sunyaev-Zeldovich effect.  In Sazonov \& Sunyaev (1998b), a numerical table which summarizes the results of the Monte Carlo calculations has been presented.  This table is of great value when one wishes to calculate the relativistic corrections to the Sunyaev-Zeldovich effect for galaxy clusters of extremely high temperatures.  Accurate analytic fitting formulae would be still more convenient to use for the observers who wish to analyze the galaxy clusters with extremely high temperatures.  This project has been undertaken by Nozawa et al. (2000).  They have presented an accurate analyic fitting formula that has a high accuracy for the ranges $0.00 \leq \theta_{e} \leq 0.05$ and $0 \leq X \leq 20$, where $\theta_{e} \equiv k_{B}T_{e}/m_{e}c^{2}$, $X \equiv \hbar \omega/k_{B}T_{0}$.  Another fitting formula with a still higher precision that is valid for the more limited ranges $0.00 \leq \theta_{e} \leq 0.035$, $0 \leq X \leq 15$ has been invented by Itoh et al. (2002b).  Relativistic corrections to the double scattering effect on the Sunyaev-Zeldovich effect have been calculated by Itoh et al. (2001).  They have presented the Fokker-Planck expansion results up to the $\theta_{e}^{6}$ correction term as well as the exact numerical results that have been obtained by the numerical integration of the Boltzmann collision term.  Dolgov et al. (2001) have carried out an independent numerical calculation.  Their result shows an excellent agreement with that of Itoh et al. (2001) for the case of small optical depth.

  It now appears that all the necessary theoretical tools are ready for the accurate analysis of the observational data of the Sunyaev-Zeldovich effect for clusters of galaxies.  However, we notice that the numerical data of Sazonov \& Sunyaev (1998b) and the fitting formula of Nozawa et al. (2000) are both valid for the range $\theta_{e} \leq 0.05$.  The recent advances in X-ray and Sunyaev-Zeldovich effect observations of galaxy clusters are likely to discover galaxy clusters with extremely high electron temperatures $k_{B}T_{e} \geq 25$keV.  Therefore it is important to present accurate numerical data for this temperature regime.  This is the task to which we will address ourselves in this paper.  For the analyses of the galaxy clusters with extremely high temperatures, the results of the calculation of the relativistic thermal bremsstrahlung Gaunt factor (Nozawa, Itoh, \& Kohyama 1998a) and their accurate analytic fitting formulae (Itoh et al. 2000b) will be useful.  The nonrelativistic electron-electron thermal bremsstrahlung Gaunt factor has been also calculated by Itoh et al. (2002a).

  The present paper is organized as follows.  In $\S$ 2 we give the method of the calculation.  In $\S$ 3 we give the numerical table.  In $\S$ 4 we give the analytic fitting formula.  Concluding remarks will be given in $\S$ 5.

\section{BOLTZMANN EQUATION}

  We will formulate the kinetic equation for the photon distribution function using a relativistically covariant formalism (Berestetskii, Lifshitz, \& Pitaevskii 1982; Buchler \& Yueh 1976).  As a reference system, we choose the system which is fixed to the center of mass of the cluster of galaxies.  This choice of the reference system affords us to carry out all the calculations in the most straightforward way.  We will use the invariant amplitude for the Compton scattering as given by Berestetskii, Lifshitz, \& Pitaevskii (1982) and by Buchler \& Yueh (1976).

 The time evolution of the photon distribution function $n(\omega)$ is written as 
\begin{eqnarray}
\frac{\partial n(\omega)}{\partial t} & = & -2 \int \frac{d^{3}p}{(2\pi)^{3}} d^{3}p^{\prime} d^{3}k^{\prime} \, W \,
\left\{ n(\omega)[1 + n(\omega^{\prime})] f(E) - n(\omega^{\prime})[1 + n(\omega)] f(E^{\prime}) \right\} \, ,  \\
W & = & \frac{(e^{2}/4\pi)^{2} \, \overline{X} \, \delta^{4}(p+k-p^{\prime}-k^{\prime})}{2 \omega \omega^{\prime} E E^{\prime}} \, ,  \\
\overline{X} & = & - \left( \frac{\kappa}{\kappa^{\prime}} + \frac{\kappa^{\prime}}{\kappa} \right) + 4 m^{4} \left( \frac{1}{\kappa} + \frac{1}{\kappa^{\prime}} \right)^{2} 
 - 4 m^{2} \left( \frac{1}{\kappa} + \frac{1}{\kappa^{\prime}} \right) \, ,  \\
\kappa & = & - 2 (p \cdot k) \, = \, - 2 \omega E \left( 1 - \frac{\mid \vec{p} \mid}{E} {\rm cos} \, \alpha \right) \, ,  \\
\kappa^{\prime} & = &  2 (p \cdot k^{\prime}) \, = \, 2 \omega^{\prime} E \left( 1 - \frac{\mid \vec{p} \mid}{E} {\rm cos} \, \alpha^{\prime} \right) \, .
\end{eqnarray}
In the above $W$ is the transition probability corresponding to the Compton scattering.  The four-momenta of the initial electron and photon are $p = (E, \vec{p})$ and $k = (\omega, \vec{k})$, respectively.  The four-momenta of the final electron and photon are $p^{\prime} = (E^{\prime}, \vec{p}^{\prime})$ and $k^{\prime} = (\omega^{\prime}, \vec{k}^{\prime})$, respectively.  The angles $\alpha$ and $\alpha^{\prime}$ are the angles between $\vec{p}$ and $\vec{k}$, and between $\vec{p}$ and $\vec{k}^{\prime}$, respectively.  Throughout this paper, we use the natural unit $\hbar = c = 1$ unit, unless otherwise stated explicitly.

  By ignoring the degeneracy effects, we have the relativistic Maxwellian distribution for electrons with temperature $T_{e}$ as follows
\begin{eqnarray}
f(E) & = & \left[ e^{\left[(E - m)-(\mu - m) \right]/k_{B}T_{e}} \, + \, 1 \right]^{-1}  \nonumber \\
& \approx & e^{-\left[K-(\mu - m)\right]/k_{B}T_{e}} \, ,
\end{eqnarray}
where $K \equiv (E - m)$ is the kinetic energy of the initial electron, and $(\mu - m)$ is the non-relativistic chemical potential of the electron. 
We now introduce the quantities
\begin{eqnarray}
x &  \equiv &  \frac{\omega}{k_{B}T_{e}}  \, ,  \\
\Delta x &  \equiv &  \frac{\omega^{\prime} - \omega}{k_{B}T_{e}}  \, .
\end{eqnarray}
Substituting equations (2.6) -- (2.8) into equation (2.1), we obtain
\begin{equation}
\frac{\partial n(\omega)}{\partial t} =  -2 \int \frac{d^{3}p}{(2\pi)^{3}} d^{3}p^{\prime} d^{3}k^{\prime} \, W \, f(E) \,
\left\{ [1 + n(\omega^{\prime})] n(\omega) -  [1 + n(\omega)] n(\omega^{\prime}) e^{ \Delta x }  \right\} \, .
\end{equation}
Equation (2.9) is our basic equation.  We will denote the Thomson scattering cross section by $\sigma_{T}$, and the electron number density by $N_{e}$.  We will define
\begin{eqnarray}
\theta_{e} & \equiv & \frac{k_{B}T_{e}}{m_{e}c^{2}}  \, ,  \\
y & \equiv & \sigma_{T} \int d \ell N_{e} \, ,
\end{eqnarray}
where $T_{e}$ is the electron temperature, and the integral in equation (2.11) is over the path length of the galaxy cluster.  By introducing the initial photon distribution of the CMB radiation which is assumed to be Planckian with temperature $T_{0}$
\begin{eqnarray}
n_{0} (X) & = & \frac{1}{e^{X} - 1} \, ,   \\
X & \equiv & \frac{\omega}{k_{B} T_{0}}  \, ,
\end{eqnarray}
we rewrite equation (2.9) as
\begin{equation}
\frac{\Delta n(X)}{n_{0}(X)} \, = \, y \, F(\theta_{e}, X)  \, .
\end{equation}
We obtain the function $F(\theta_{e}, X)$ by numerical integration of the collision term of the Boltzmann equation (2.9).  The accuracy of the numerical integration is about $10^{-5}$.  We confirm that the condition of the photon number conservation
\begin{equation}
\int d X \, X^{2} \, \Delta n(X) \, = \, 0  \,
\end{equation}
is satisfied with the accuracy better than $10^{-9}$.

  We define the distortion of the spectral intensity as
\begin{eqnarray}
\Delta I & \equiv & \frac{X^{3}}{e^{X} - 1} \, \frac{\Delta n(X)}{n_{0}(X)}  \,     \\
& = &  y \, \frac{X^{3}}{e^{X} - 1} \, F(\theta_{e}, X)  \, .
\end{eqnarray}
The graph of $\Delta I/y$ is shown in Figure 1.

\section{NUMERICAL RESULTS}

  In Tables 1--5 we give the numerical results of the function
\begin{eqnarray}
\frac{\Delta I}{y} & = & \frac{1}{y} \frac{X^{3}}{e^{X} - 1} \, \frac{\Delta n(X)}{n_{0}(X)}  \,  \nonumber \\
& = &  \frac{X^{3}}{e^{X} - 1} \, F(\theta_{e}, X)  \, .
\end{eqnarray}

\section{ANALYTIC FITTING FORMULA}

  We express the analytic fitting formula for $0.05 \leq \theta_{e} \leq 0.10$, $0 \leq X \leq 17$ in the following way.

\noindent
(i) For $0 \leq X \leq 1.2$
\begin{eqnarray}
\frac{\Delta I}{y} & = & \frac{1}{y} \frac{X^{3}}{e^{X} - 1} \, \frac{\Delta n(X)}{n_{0}(X)}  \,    \nonumber  \\
  &  =  &  \frac{\theta_{e} X^{4} e^{X}}{ \left(e^{X}-1 \right)^{2}} \, \left(
Y_{0} \, + \, \theta_{e} Y_{1} \, + \, \theta_{e}^{2} Y_{2} \, + \,  \theta_{e}^{3} Y_{3} \, + \,  \theta_{e}^{4} Y_{4} \right) 
\end{eqnarray}
\begin{eqnarray}
Y_{0} & = & - 4 \, + \tilde{X}  \,  , \\
Y_{1} & = & - 10 + \frac{47}{2} \tilde{X} - \frac{42}{5} \tilde{X}^{2} + \frac{7}{10} \tilde{X}^{3}  \, + \, \tilde{S}^{2} \left( - \frac{21}{5} + \frac{7}{5} \tilde{X} \right) \,  ,  \\
Y_{2} & = & - \frac{15}{2} + \frac{1023}{8} \tilde{X} - \frac{868}{5} \tilde{X}^{2} + \frac{329}{5} \tilde{X}^{3} - \frac{44}{5} \tilde{X}^{4} + \frac{11}{30} \tilde{X}^{5}  \nonumber \\ 
& + & \tilde{S}^{2} \left( - \frac{434}{5} + \frac{658}{5} \tilde{X}  - \frac{242}{5}  \tilde{X}^{2} + \frac{143}{30} \tilde{X}^{3} \right)  \nonumber  \\
& + &  \tilde{S}^{4} \left( - \frac{44}{5} + \frac{187}{60} \tilde{X} \right) \, , \\
Y_{3} & = & \frac{15}{2} + \frac{2505}{8} \tilde{X} - \frac{7098}{5} \tilde{X}^{2} + \frac{14253}{10} \tilde{X}^{3} - \frac{18594}{35} \tilde{X}^{4}   \nonumber  \\
& + &  \frac{12059}{140} \tilde{X}^{5} - \frac{128}{21} \tilde{X}^{6} + \frac{16}{105} \tilde{X}^{7} \nonumber \\ 
& + & \tilde{S}^{2} \left( - \frac{7098}{10} + \frac{14253}{5} \tilde{X} - \frac{102267}{35}  \tilde{X}^{2} + \frac{156767}{140} \tilde{X}^{3} - \frac{1216}{7}  \tilde{X}^{4} + \frac{64}{7} \tilde{X}^{5} \right)  \nonumber  \\
& + &  \tilde{S}^{4} \left( - \frac{18594}{35} + \frac{205003}{280} \tilde{X} - \frac{1920}{7}  \tilde{X}^{2} + \frac{1024}{35} \tilde{X}^{3} \right) \nonumber  \\
& + &  \tilde{S}^{6} \left( - \frac{544}{21} + \frac{992}{105} \tilde{X} \right) \, , \\
Y_{4} & = & - \frac{135}{32} + \frac{30375}{128} \tilde{X} - \frac{62391}{10} \tilde{X}^{2} + \frac{614727}{40} \tilde{X}^{3} - \frac{124389}{10} \tilde{X}^{4}   \nonumber  \\
& + &  \frac{355703}{80} \tilde{X}^{5} - \frac{16568}{21} \tilde{X}^{6} + \frac{7516}{105} \tilde{X}^{7} - \frac{22}{7} \tilde{X}^{8} + \frac{11}{210} \tilde{X}^{9} \nonumber \\ 
& + & \tilde{S}^{2} \left( - \frac{62391}{20} + \frac{614727}{20} \tilde{X} - \frac{1368279}{20} \tilde{X}^{2} + \frac{4624139}{80} \tilde{X}^{3} - \frac{157396}{7}  \tilde{X}^{4}  \right. \nonumber  \\
&  & \, \, \, \, \, + \, \left. \frac{30064}{7} \tilde{X}^{5} - \frac{2717}{7} \tilde{X}^{6} + \frac{2761}{210} \tilde{X}^{7}   \right)  \nonumber  \\
& + &  \tilde{S}^{4} \left( - \frac{124389}{10} + \frac{6046951}{160} \tilde{X} - \frac{248520}{7} \tilde{X}^{2} + \frac{481024}{35} \tilde{X}^{3} - \frac{15972}{7} \tilde{X}^{4}  \right. \nonumber  \\
&  &  \, \, \, \, + \, \left. \frac{18689}{140} \tilde{X}^{5}  \right) \nonumber  \\
& + &  \tilde{S}^{6} \left( - \frac{70414}{21} + \frac{465992}{105} \tilde{X} - \frac{11792}{7} \tilde{X}^{2} + \frac{19778}{105} \tilde{X}^{3} \right) \nonumber  \\
& + &  \tilde{S}^{8} \left( - \frac{682}{7} + \frac{7601}{210} \tilde{X} \right) \, , 
\end{eqnarray}
where
\begin{eqnarray}
\tilde{X} & \equiv &  X \, {\rm coth} \left( \frac{X}{2} \right)  \, , \\
\tilde{S} & \equiv & \frac{X}{ \displaystyle{ {\rm sinh} \left( \frac{X}{2} \right)} }   \, .
\end{eqnarray}

\noindent
(ii) For $1.2 \leq X \leq 17$
\begin{eqnarray}
\frac{\Delta I}{y} &  =  &  \theta_{e} \, X^{2}\,  e^{-X} \, \left(X - X_{0} \right) \, G( \theta_{e}, X) \, , 
\end{eqnarray}
In equation (4.9) $X_{0}$ is the crossover frequency, where the thermal Sunyaev-Zeldovich effect vanishes.  We use the following expression for $X_{0}$ obtained by Itoh et al. (1998).
\begin{eqnarray}
X_{0} & = & 3.830 (1 + 1.1674 \theta_{e} - 0.8533 \theta_{e}^{2} ) \, ,
\end{eqnarray}
The explicit form of the analytic fitting function $G(\theta_{e}, X)$ is given  as follows:
\begin{eqnarray}
G(\theta_{e}, X) & = & \sum_{i,j=0}^{12} \, a_{i \, j} \, \Theta_{e}^{i} \, Z^{j} \, ,  \\
\Theta_{e} & \equiv & 10 \, \theta_{e} \, , \, \, \, 0.05 \leq \theta_{e} \leq 0.100  \, ,  \\
Z  & \equiv &  \frac{1}{20} X \, , \, \, \, 1.2 \leq X \leq 17 \, .
\end{eqnarray}
The coefficients $a_{i \, j}$ are presented in Table 6.  The accuracy of the fitting formula is generally better than 0.4\% except for the crossover frequency region $X \approx 4.0$ where the thermal Sunyaev-Zeldovich signal almost exactly vanishes.  The accuracy of the presented fitting formula is shown in Figure 2.

\section{CONCLUDING REMARKS}

We have calculated the relativistic corrections to the thermal Sunyaev-Zeldovich effect for clusters of galaxies by numerical integration of the collision term of the Boltzmann equation.  The emphasis is placed on the cases of high electron temperatures in order to provide the theoretical tool for the analysis of extremely high-temperature galaxy clusters with $k_{B}T_{e} \geq 25$keV that are likely to be discovered by the ongoing X-ray and Sunyaev-Zeldovich effect observations.  We have presented a detailed numerical table and an accurate analytic fitting formula.  We will be most pleased to provide the computer subroutines for the numerical table with a standard interpolation package as well as for the analytic fitting formula if we are requested to do so.  We sincerely wish the present results together with our previous results would contribute to the accurate analysis of the Sunyaev-Zeldovich effect observations of galaxy clusters that are likely to produce the most interesting results on the formation and evolution of galaxy clusters.

\acknowledgements

  We thank Y. Oyanagi for allowing us to use the least square fitting program SALS.  We also thank Y. Suto and T. Kitayama for discussions on their observational results that have inspired us to write this paper.  It is also a pleasure to thank Y. Rephaeli for the discussion on the calculation of the thermal Sunyaev-Zeldovich effect for high-temperature galaxy clusters.  This work is financially supported in part by the Grant-in-Aid of Japanese Ministry of Education, Culture, Sports, Science, and Technology under the contract \#13640245.

\newpage


\references{}
\reference{} Allen, S. W., Ettori, S., \& Fabian, A. C. 2001, MNRAS, 324, 877
\reference{} Arnaud, K. A., Mushotzky, R. F., Ezawa, H., Fukazawa, Y., Ohashi, T., Bautz, M. W., Crewe, G. B., Gendreau, K. C., Yamashita, K., Kamata, Y., \& Akimoto, F. 1994, ApJ, 436, L67
\reference{} Audit, E., \& Simmons, J. F. L. 1999, MNRAS, 305, L27
\reference{} Berestetskii, V. B., Lifshitz, E. M., \& Pitaevskii, L. P. 1982, $Quantum$ $Electrodynamics$ (Oxford: Pergamon)
\reference{} Birkinshaw, M. 1999, Physics Reports, 310, 97
\reference{} Buchler, J. R., \& Yueh, W. R. 1976, ApJ, 210, 440
\reference{} Carlstrom, J. E., Holder, G. P., \& Reese, E. D. 2002, ARA\&A, 40, 643
\reference{} Challinor, A., Ford, M., \& Lasenby, A., 1999, MNRAS, 312, 159.
\reference{} Challinor, A., \& Lasenby, A., 1998, ApJ, 499, 1
\reference{} Challinor, A., \& Lasenby, A., 1999, ApJ, 510, 930
\reference{} David, L. P., Slyz, A., Jones, C., Forman, W., \& Vrtilek, S. D. 1993, ApJ, 412, 479
\reference{} Dolgov, A. D., Hansen, S. H., Pastor, S., \& Semikoz, D. V. 2001, ApJ, 554, 74
\reference{} Hansen, E., \& Lilje, P. B. 1999, MNRAS, 306, 153
\reference{} Holzapfel, W. L. et al. 1997, ApJ, 480, 449
\reference{} Itoh, N., Kawana, Y., \& Nozawa, S. 2002a, Nuovo Cimento, 117B, 359\reference{} Itoh, N., Kawana, Y., Nozawa, S., \& Kohyama, Y. 2001, MNRAS, 327, 567
\reference{} Itoh, N., Kohyama, Y., \& Nozawa, S. 1998, ApJ, 502, 7
\reference{} Itoh, N., Nozawa, S., \& Kohyama, Y. 2000a, ApJ, 533, 588
\reference{} Itoh, N., Sakamoto, T., Kusano, S., Nozawa, S., \& Kohyama, Y. 2000b, ApJS, 128, 125
\reference{} Itoh, N., Sakamoto, T., Kusano, S., Kawana, Y., \& Nozawa, S. 2002b, A\&A, 382, 722
\reference{} Kompaneets, A. S. 1957, Soviet Physics JETP, 4, 730
\reference{} Markevitch, M. 1998, ApJ, 504, 27
\reference{} Markevitch, M., Mushotzky, R., Inoue, H., Yamashita, K., Furuzawa, A., \& Tawara, Y. 1996, ApJ, 456, 437
\reference{} Markevitch, M., Yamashita, K., Furuzawa, A., \& Tawara, Y. 1994, ApJ, 436, L71
\reference{} Mushotzky, R. F., \& Scharf, C. A. 1997, ApJ, 482, L13
\reference{} Nozawa, S., Itoh, N., Kawana, Y., \& Kohyama, Y. 2000, ApJ, 536, 31\reference{} Nozawa, S., Itoh, N., \& Kohyama, Y. 1998a, ApJ, 507, 530
\reference{} Nozawa, S., Itoh, N., \& Kohyama, Y. 1998b, ApJ, 508, 17
\reference{} Rephaeli, Y. 1995, ApJ, 445, 33
\reference{} Rephaeli. Y., \& Yankovitch, D. 1997, ApJ, 481, L55
\reference{} Sazonov, S. Y., \& Sunyaev, R. A. 1998a, ApJ, 508, 1
\reference{} Sazonov, S. Y., \& Sunyaev, R. A. 1998b, Astronomy Letters 24, 553
\reference{} Sazonov, S. Y., \& Sunyaev, R. A. 1999, MNRAS, 310, 765
\reference{} Schmidt, R. W., Allen, S. W., \& Fabian, A. C. 2001, MNRAS, 327, 1057
\reference{} Stebbins, A., 1997, preprint astro-ph/9705178
\reference{} Sunyaev, R. A., \& Zeldovich, Ya. B. 1972, Comments Astrophys. Space Sci., 4, 173
\reference{} Sunyaev, R. A., \& Zeldovich, Ya. B. 1980a, ARA\&A, 18, 537
\reference{} Sunyaev, R. A., \& Zeldovich, Ya. B. 1980b, MNRAS, 190, 413
\reference{} Sunyaev, R. A., \& Zeldovich, Ya. B. 1981, Astrophys. Space Phys. Rev., 1, 1
\reference{} Tucker, W., Blanco, P., Rappoport, S., David, L., Fabricant, D., Falco, E. E., Forman, W., Dressler, A., \& Ramella, M. 1998, ApJ, 496, L5
\reference{} Weymann, R. 1965, Phys. Fluid, 8, 2112
\reference{} Zeldovich, Ya. B., \& Sunyaev, R. A. 1969, Ap\&SS, 4, 301


\newpage
\centerline{\bf \large Figure Legend}

\begin{itemize}

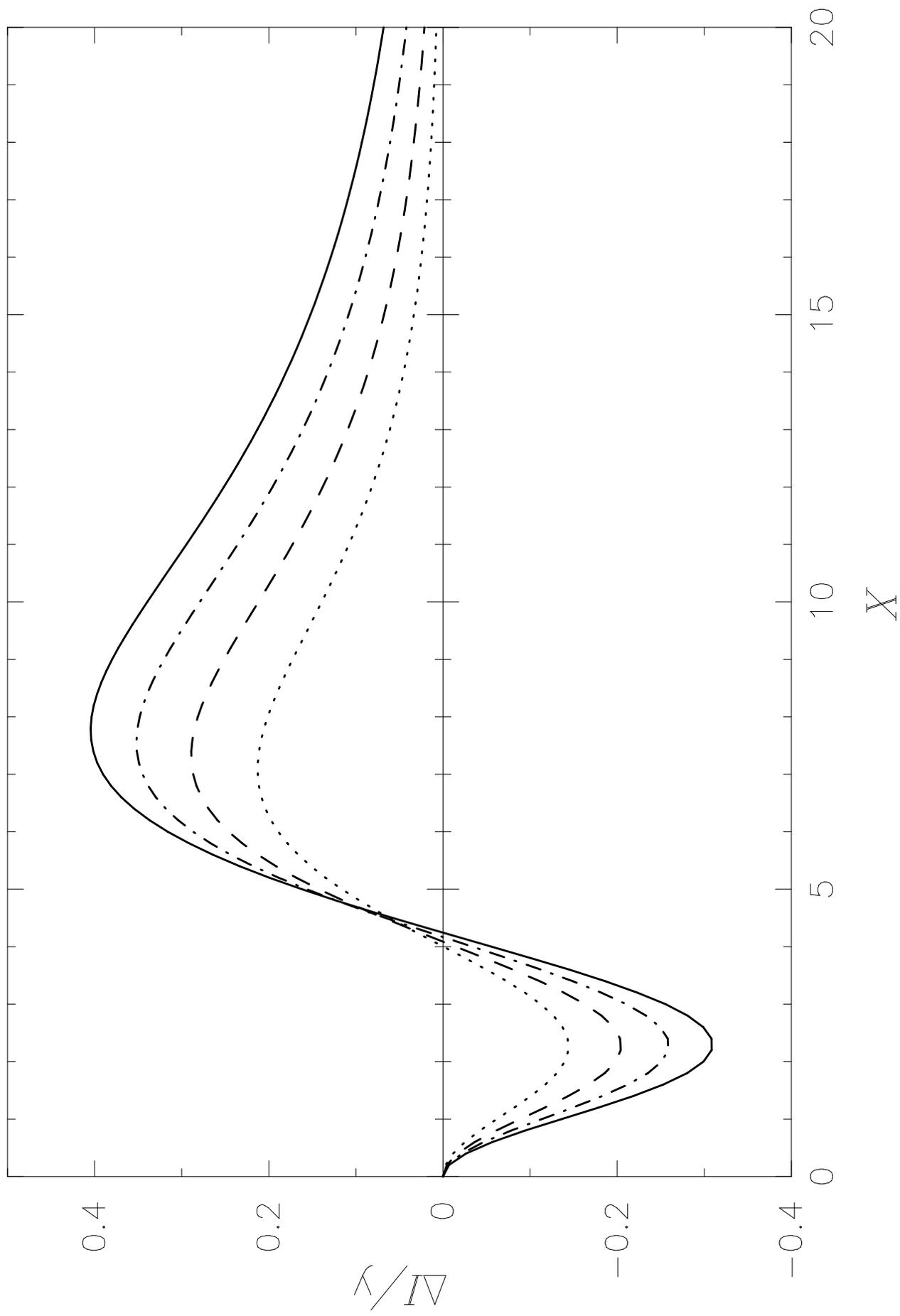
\item Figure 1. Spectral intensity distortion $\Delta I/y$ as a function of $X$.  The dotted, dashed, dash-dotted, and solid curves correspond to the cases for $\theta_{e}=0.04, 0.06, 0.08, 0.10$, respectively.

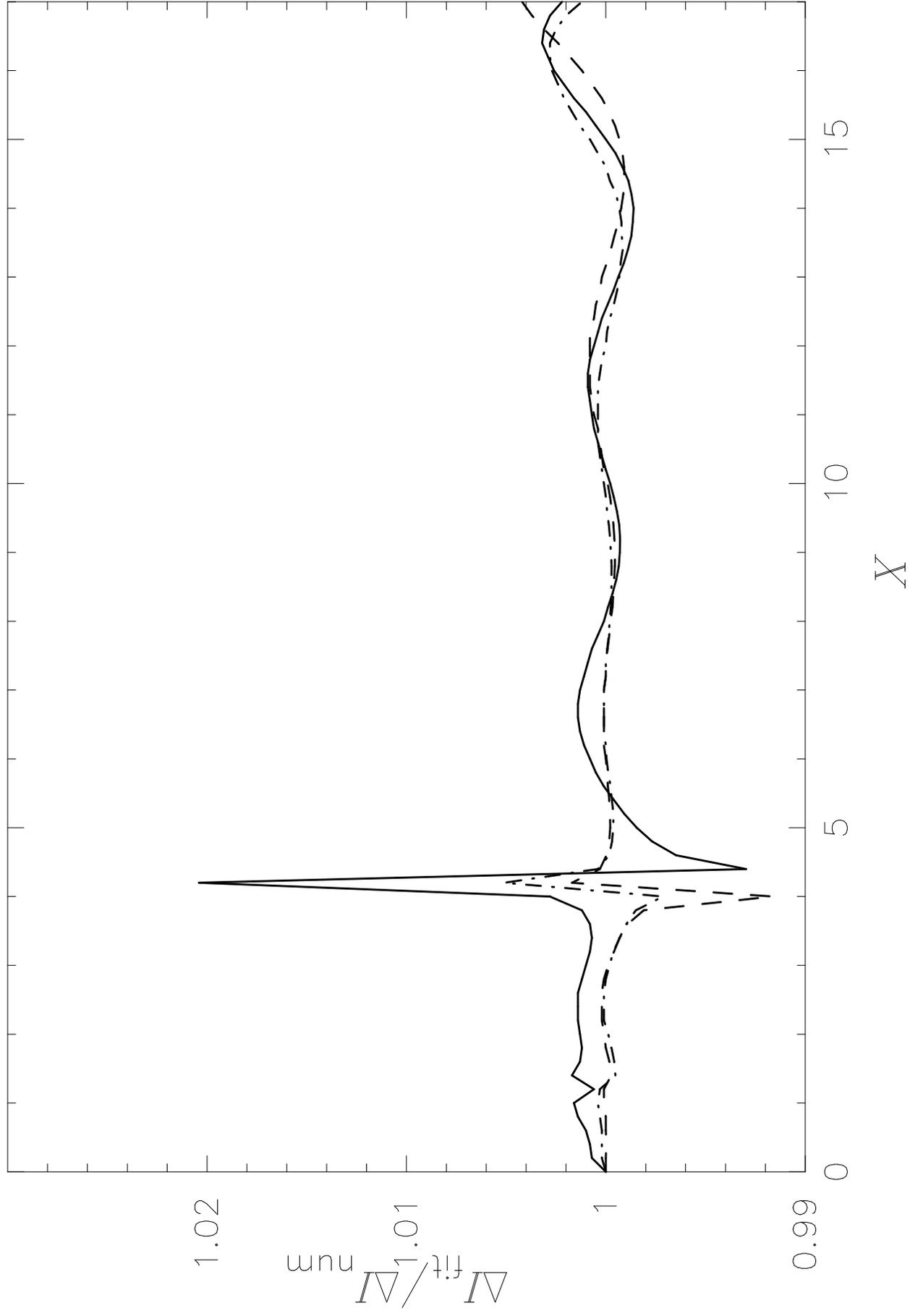
\item Figure 2.  The accuracy of the present fitting formula for various values of  $\theta_{e}$.  The dashed, dash-dotted, and solid curves correspond to  $\theta_{e}$=0.050, 0.075, and 0.100, respectively.
\end{itemize}

\end{document}